\documentclass[pra,aps,floatfix,showpacs,tightenlines,superscriptaddress,amsmath,amssymb,showkeys,10pt,nofootinbib]{revtex4-2}
\usepackage{slashed}
\usepackage{xcolor}
\usepackage{graphicx}

\usepackage{hyperref}

\newcommand{\be}{\begin{equation}}\newcommand{\ee}{\end{equation}}
\newcommand{\bea}{\begin{eqnarray}}\newcommand{\eea}{\end{eqnarray}}
\newcommand{\brr}{\begin{array}}\newcommand{\err}{\end{array}}
\newcommand{\bit}{\begin{itemize}}\newcommand{\eit}{\end{itemize}}
\newcommand{\ben}{\begin{enumerate}}\newcommand{\een}{\end{enumerate}}

\newcommand{\bbm}{\begin{bmatrix}}\newcommand{\ebm}{\end{bmatrix}}
\newcommand{\ba}{\begin{array}}
\newcommand{\ea}{\end{array}}
\newcommand{\G}{\textbf}

\newtheorem{mydef}{Definition}
\newtheorem{Lemma}{Lemma}
\newcommand{\bd}{\begin{mydef}} \newcommand{\ed}{\end{mydef}}
\newcommand{\bthe}{\begin{theorem}} \newcommand{\ethe}{\end{theorem}}
\newcommand{\ble}{\begin{Lemma}} \newcommand{\ele}{\end{Lemma}}

\newcommand{\dr}{\mathrm{d}}

\def\ha{\frac{1}{2}}

\def\intx{\int \!\!\mathrm{d}^3 {\G x}}

\def\lan{\langle}
\def\lf{\left}

\def\non{\nonumber}\def\pa{\partial}\def\ran{\rangle}

\def\ri{\right}
\def\al{\alpha}\def\bt{\beta}\def\ga{\gamma}
\def\de{\delta}\def\De{\Delta}

\def\si{\sigma}
\def\om{\omega}

\def\1{{_{1}}}\def\2{{_{2}}}

\newcommand{\ide}{1\hspace{-1mm}{\rm I}}

\def\noHe0{:\;\!\!\;\!\!:H_L(0):\;\!\!\;\!\!:}
\def\noHm0{:\;\!\!\;\!\!:H_\mu(0):\;\!\!\;\!\!:}

\def\lan{\langle}
\def\lf{\left}

\def\non{\nonumber}
\def\pa{\partial}\def\ran{\rangle}

\def\ri{\right}

\def\al{\alpha}\def\bt{\beta}\def\ga{\gamma}
\def\de{\delta}\def\De{\Delta}

\def\si{\sigma}
\def\om{\omega}

\def\1{{_{1}}}\def\2{{_{2}}}

\begin{document}

\title{Chiral oscillations in finite time quantum field theory}

\author{Massimo Blasone}
\email{blasone@sa.infn.it}
\affiliation{Dipartimento di Fisica, Universit\`a di Salerno, Via Giovanni Paolo II 132, 84084 Fisciano (SA), Italy}
\affiliation{INFN Sezione di Napoli, Gruppo collegato di Salerno, Italy}

\author{Francesco Giacosa}
\email{fgiacosa@ujk.edu.pl}
\affiliation{Institute of Physics, Jan-Kochanowski University, ul. Uniwersytecka 7, 25-406 Kielce, Poland}
\affiliation{Institute for Theoretical Physics, J. W. Goethe University, Max-von-Laue-Straße 1,
60438 Frankfurt, Germany}

\author{Luca Smaldone}
\email{lsmaldone@unisa.it}
\affiliation{Dipartimento di Fisica, Universit\`a di Salerno, Via Giovanni Paolo II 132, 84084 Fisciano (SA), Italy}
\affiliation{INFN Sezione di Napoli, Gruppo collegato di Salerno, Italy}

\author{Giorgio Torrieri}
\email{torrieri@unicamp.br}
\affiliation{Instituto de Fisica Gleb Wataghin - UNICAMP, 13083-859, Campinas SP, Brazil}
\affiliation{Institute of Physics, Jan-Kochanowski University, ul. Uniwersytecka 7, 25-406 Kielce, Poland}

\begin{abstract}
We demonstrate how chiral oscillations of a massive Dirac field can be described within quantum field theory using a finite-time interaction picture approach, where the mass term in the Lagrangian is treated as a perturbative coupling between massless fields of definite chirality. We derive the formula for chiral oscillations at the fourth order in the perturbative expansion, obtaining a result consistent with the formula derived by means of other methods. 
Furthermore, we illustrate how the perturbative framework of chiral oscillations can effectively describe production processes where an electron must exhibit both left chirality and positive helicity, as in decay $\pi^- \to e^- + {\bar \nu}_e$. Finally, we argue that, in this perturbative view, chiral oscillations are also essential for detecting the decay products in such processes.
\end{abstract}

\maketitle
%%%%%%%%%%%%%%%%%%%%%%%%%%%%%%%%%%%%%%%%%%%%%%%%%%%%%%%%%%%%%%%%%%%%%%%%%%%%%%%
\section{Introduction}

One of the key aspects of the Standard Model is that charged current weak interactions involve fermions with definite chirality \cite{PhysRevLett.19.1264,Salam:1968rm,pal2014introductory}: only left-chiral particles states and right-chiral antiparticles state participate in weak interactions. However,  chirality of massive particles is not preserved during time evolution: consequently particles produced in weak interactions undergo chiral oscillations \cite{DeLeo:1996gt,fukugita2003physics,Bernardini:2005wh,Bernardini:2005cu,Bernardini:2006bq,Bernardini:2006cy,Bernardini:2006cn,Bernardini:2007ew,Bernardini:2007uf,Bittencourt:2020xen,suekane2021quantum,Bittencourt:2022hwn}. Originally, chiral oscillations were proposed to explain the solar neutrino puzzle \cite{DeLeo:1996gt}. Today, we know that the reduced flux of electron neutrinos from the Sun is caused by neutrino flavor oscillations \cite{fukugita2003physics,giunti2007fundamentals,Smaldone:2021mii}. However, chiral oscillations still hold some phenomenological interest. In fact, it has been recently suggested that this phenomenon could be observable in the cosmic neutrino background \cite{Ge:2020aen,Bittencourt:2020xen} and in electronic transport in graphene layers where chiral symmetry is explicitly broken by external potential barriers \cite{PhysRevB.102.205404}.

Recently, it has been proposed \cite{Blasone:2023brf,Blasone:2024zsn} that neutrino flavor oscillation processes can be described using the usual perturbation series in the interaction picture, where the mixing among different flavors acts as an interaction. In this framework, where calculations at finite times are involved, the use of the time-evolution operator is essential because the $S$-matrix would disrupt flavor oscillations. This is due to the time-energy uncertainty relation, which sets a lower bound on energy resolution in flavor oscillations \cite{Bilenky2007,Bilenky2008,Blasone2019,Blasone2020}. Notably, the interaction picture computation, within the approximation used for truncating the perturbative expansion, leads to the same oscillation formula originally derived within a non-perturbative quantum field theoretical (QFT) approach \cite{Blasone:1995zc,BHV99}.

An approach to chiral oscillations in QFT has been proposed in Ref. \cite{Bittencourt:2024yxi} where chiral charges are diagonalized by means of a suitable Bogoliubov transformation. There, it has been proved that the Hilbert space for chiral states is unitarily inequivalent to the usual Fock space for the energy eigenstates. The chiral oscillation probability is then computed as the expectation value of the chiral charge on a state with definite chirality. The result coincides with the usual chiral oscillations probability calculated in relativistic quantum mechanics.

In this paper, we investigate how chiral oscillations of a massive Dirac field can be described within a plain and well-defined QFT approach using a standard perturbative expansion and finite-time intervals, where the mass term in the Lagrangian is treated as an interaction between massless fields with definite chiralities. In particular, this method does not rely on Bogoliubov transformations but resembles the study of unstable states in QFT \cite{Giacosa:2011xa,Giacosa:2021hgl}. Remarkably, our results coincide with the conventional quantum-mechanical predictions (and consequently with Ref. \cite{Bittencourt:2024yxi}), within the approximation framework adopted in the interaction picture. 
Additionally, we demonstrate how the perturbative framework for chiral oscillations can be used to describe production processes where an electron must simultaneously exhibit left chirality and positive helicity, as in the decay $\pi^- \to e^- + {\bar \nu}_e$. Lastly, we argue that, within this perturbative perspective, chiral oscillations are also crucial for detecting decay products in such processes.

%%%%%%%%%%%%%%%%%%%%%%%%%%%%%%%%%%%%%%%%%%%%%%%%%%%%%%%%%%%%%%%%%%%%%%%%%%%%%%%%%%%%%%%%%%%%%%%%%%%%
\section{Chiral oscillations} \label{comp}

Here we briefly review chiral oscillations in relativistic quantum mechanics. The Dirac equation in the Hamiltonian form reads
\be
H_D |\psi(t) \ran \ = \ \lf( \boldsymbol{\al} \cdot \G P+ \bt \, m\ri)|\psi(t)\ran \ = \ i \pa_t |\psi(t)\ran \, ,
\ee
where $\G P$ is the three-momentum operator and $\boldsymbol{\al} \equiv (\al_1,\al_2,\al_3)$, $\bt$ are the Dirac matrices. 

The spinors with a definite chirality are defined as
\be
|\psi_{L (R)}(t)\ran \ = \ N \, P_{L(R)} \, |\psi(t)\ran \, , 
\ee
where we introduced the usual chiral projectors $P_L(R) \equiv \frac{1-(+)\ga_5}{2}$, with $\ga_5 = i \al_1 \al_2 \al_3$ being the chiral matrix, and where $N$ is a normalization factor. In the chiral representation $\ga_5 = diag(-\ide_2,\ide_2)$ so that $|\psi\ran=(|\xi_L,\xi_R\ran)$, with $|\psi_R\ran=(0,0,|\xi_R\ran)$, $|\psi_L\ran=(|\xi_L\ran,0,0)$. The two component spinors satisfy the coupled equations
\bea
i \pa_t |\xi_L(t)\ran & = & \boldsymbol{\si} \cdot \G P  |\xi_L(t)\ran + m  |\xi_R(t)\ran \, , \\[2mm]
i \pa_t |\xi_R(t)\ran & = & \boldsymbol{\si} \cdot \G P  |\xi_R(t)\ran + m  |\xi_L(t)\ran \, ,
\eea
where $\boldsymbol{\si} \equiv (\si_1,\si_2,\si_3)$ are the Pauli matrices. Notice that the mass couples the two equations. Otherwise, these would be two decoupled \emph{Weyl equations}. For such reason chirality is not conserved. In other words a massive fermion will undergo a \emph{chiral oscillation} during its propagation.

It is well-known that charged current weak interactions produce left (right)-chiral (anti)fermions. Then suppose that a left-chiral fermion is produced at $t=t_i$.
The survival probability will thus be given by (for a derivation see Refs.
 \cite{DeLeo:1996gt,Bernardini:2005wh,Bernardini:2005cu,Bittencourt:2020xen,Bittencourt:2022hwn})
\be \label{plr}
 P_{L \to L}(\G p,\De t) \ = \ |\lan \psi_L(t_f)|\psi_L(t_i)|^2 \ = \ 1-\frac{m^2}{\om_\G p^2}  \sin^2\lf(\om_\G p \, \De t\ri) \, ,  \qquad \De t \equiv t_f-t_i \, , 
\ee
with $\om_\G p = \sqrt{|\G p|^2+m^2}$ being the energy of the massive Dirac particle $\psi$. Therefore, if a fermion is produced in a weak process and then revealed through another weak process, one should observe a depletion of the produced fermions ($R$ particles are sterile) as it occurs in the case of flavor oscillations, see below.

In the following, we show how the above formula can be derived in QFT through a perturbation theory approach. 
%%%%%%%%%%%%%%%%%%%%%%%%%%%%%%%%%%%%%%%%%
\section{Chiral oscillations in the interaction picture} \label{gencon}

Let us now move on to QFT. The Lagrangian of a massive Dirac field can be written in terms of its chiral projections as
\bea 
\mathcal{L} \  =  \sum_{\si=L,R} \overline{\psi}_\si i \slashed{\pa} \psi_\si-m \lf(\overline{\psi}_L \psi_R+\overline{\psi}_R \psi_L\ri)\, , 
\label{Linteract}
\eea
We follow the idea originally developed in Ref. \cite{Blasone:2023brf} (see also \cite{Blasone:2024zsn}), where we treated mixing among neutrino flavors as an interaction within the Dirac or `interaction' picture. Here we repeat similar computations, treating the mass term of the Lagrangian \eqref{Linteract} as the interaction with respect to which we compute the perturbative expansion.
Our starting point is the Dyson formula for the time evolution operator
\be \label{dyfor}
U(t_i,t_f) \ = \ \mathcal{T} \exp \lf[i \int^{t_f}_{t_i} \!\! \dr^4 x \, :\mathcal{L}_{int}(x): \ri] \ = \ \mathcal{T}\exp  \lf[-i \int^{t_f}_{t_i} \!\! \dr^4 x \, :\mathcal{H}_{int}(x): \ri] \, ,
\ee
where $\mathcal{L}_{int} =-m \lf(\overline{\psi}_L \psi_R+\overline{\psi}_R \psi_L\ri)$, $\mathcal{H}_{int}(x)=-\mathcal{L}_{int}(x)$ is the interaction Hamiltonian density and $\mathcal{T}$ is the chronological product. Below, we only need the expression of the operator up to the second order
\be \label{dyfor2}
U(t_i,t_f) 
\ = \ 1-i\int_{t_{i}}^{t_{f}}\!\!\dr t_{1} \, : H_{int}(t_{1}):-\ha
\int_{t_{i}}^{t_{f}}\!\!\dr t_{1} 
\int_{t_{i}}^{t_{f}}\dr t_{2} \, \,: H_{int}(t_{1}) H_{int}(t_{2}): + \ldots
\ee
where $H_{int}=\intx \,\mathcal{H}_{int}(x)$ is the interaction Hamiltonian.

In the interaction picture $\psi_\si$ ($\si=L,R$) can be expanded as free fields, evolving under the action of $\mathcal{L}_0$. Such fields can be expanded as \cite{pal2014introductory}
\begin{eqnarray}
\psi_{\si}(x) = \frac{1}{\sqrt{V}} \sum_{\G k}\,  \lf( u_{{\bf k},\si}  \, \alpha_{{\bf k},\si} \,  e^{- i k x} + v_{{\bf k},\si}\, \bt_{{\bf k},\si}^{\dag} \,  e^{ i k x}   \right) \, .
\label{fieldex}
\end{eqnarray}
Note that no sum over helicities is present because we are dealing with massless fields with a definite chirality. Therefore, the index $\si$ already fixes the helicity: a left-handed fermion has negative helicity and vice-versa. In the chiral representation, left-spinors only have the first two upper components non-trivial, while the opposite is true for the right-spinors. The perturbative vacuum is defined so that
$
\al_{\G k, \si}|0 \rangle = 0 = \beta _{{\bf k},\si} |0 \rangle \  .
$
The canonical anticommutation relations are
\be  \label{CAR2} \{\al_{{\bf k},\rho}, \al ^{\dag }_{{\bf q},\si}\} = \de_{\G k \G q}\de _{\rho \si}  \quad \, , \quad \{\bt^r_{{\bf k},\rho},
\bt^{\dag }_{{\bf q},\si}\} =
\de_{\G k \G q}\de _{\rho \si} \, .
\ee
%
%In the chiral representation \cite{Miransky:1994vk}
%\be
%u_{\G k,R} \ = \ \begin{pmatrix} w_{\G k,+} \\ 0 \end{pmatrix} \, , \quad \quad v_{\G k,R} \ = \ \begin{pmatrix} \tilde{w}_{\G k,-} \\  0  \end{pmatrix}  \, \quad  u_{\G k,L} \ = \ \begin{pmatrix} 0 \\  w_{\G k,-}  \end{pmatrix} \, , \quad v_{\G k,L} \ = \ \begin{pmatrix} 0 \\ \tilde{w}_{\G k,+} \end{pmatrix} \, ,  \, , 
%\ee
%where
%\be
%w_{\G k,+} \ = \ \frac{1}{\sqrt{2 (n_3+1)}} \begin{pmatrix} n_3+1 \\ n_1+i n_2 \end{pmatrix} \, , \quad w_{\G k,-} \ = \ \frac{1}{\sqrt{2 (n_3+1)}} \begin{pmatrix}   -n_1+i n_2 \\ n_3+1 \end{pmatrix} \, , \quad n_j \equiv \frac{k_j}{|\G k|} \, , \quad \tilde{w}_{\G k,\pm} \ = \ w_{-\G k,\pm} \, .
%\ee

Using the expansion \eqref{fieldex}, the interacting Hamiltonian reads: 

\bea
H_{int}(t)& = &  m  \sum_{\G p}
 \Big[\bt_{-\G p,R}\al_{\G p, L} \, e^{-2 i |\G p| t}+\al^{\dag}_{\G p,R}\bt^{\dag}_{-\G p,L} \, e^{2 i |\G p| t}\, + \, h.c. \Big] \, .
\eea

The last ingredient are the perturbative (chiral) states which are defined in the usual way as
\be
|\psi_{\G p,\si}\ran \ \equiv \ \al^{\dag}_{\G p,\si}|0\ran \, ,
\ee

\begin{figure}[t]
\centering
\includegraphics[scale=0.6]{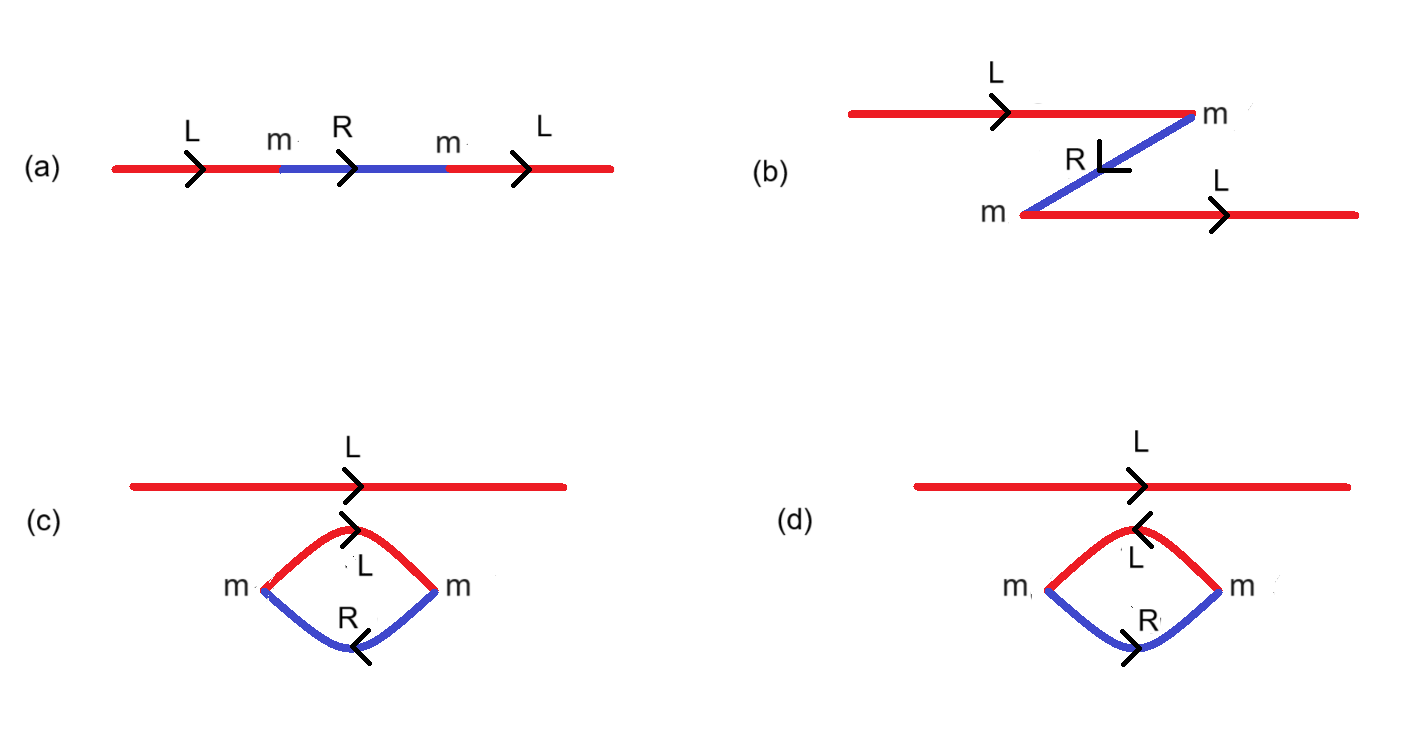}
\caption{Second-order diagrams for $L$ survival probability. Time flows from left to right. The diagram (a) vanishes since equal massless objects are involved; the Z-type diagram (b) generates a nonzero contribution. The diagrams (c) and (d) are subtracted. }
\label{fig:2th}
\end{figure}

We compute the survival probability $\mathcal{P}_{L \to L}(\G p,\De t)$ within a QFT approach involving finite time intervals, to be compared with Eq.\eqref{plr}. 
The zeroth order contribution gives  $\mathcal{P}_{L \to L}(\G k,\De t)=1$.  To find a non-trivial contribution to the survival probability we compute the second order term of Eq.\eqref{dyfor2} as:
\bea\non
&& U(t_i,t_f) \ = \ \ide -i \, m \int^{t_f}_{t_i} \!\! \dr^4 x \, : \overline{\psi}_L(x) \psi_R(x)+\overline{\psi}_R(x) \psi_L(x): \\[2mm]
&& - \  \frac{m^2}{2} \int^{t_f}_{t_i} \!\! \dr^4 x_1 \int^{t_f}_{t_i} \!\! \dr^4 x_2 \, \, \mathcal{T}\Big[   \lf(:\overline{\psi}_L(x_1) \psi_R(x_1)+\overline{\psi}_R(x_1) \psi_L(x_1):\ri) \lf( : \overline{\psi}_L(x_2) \psi_R(x_2)+\overline{\psi}_R(x_2) \psi_L(x_2):\ri)\Big] \, + \, \ldots \, . 
\eea
The second-order piece can be further expanded using Wick's theorem:
\bea \non
U^{(2)}(t_i,t_f) & = & -\frac{m^2}{2} \int^{t_f}_{t_i} \!\! \dr^4 x_1 \int^{t_f}_{t_i} \!\! \dr^4 x_2 \, \,    \Big[:\overline{\psi}_L(x_1) \psi_R(x_1)\overline{\psi}_L(x_2) \psi_R(x_2):+:\overline{\psi}_L(x_1) \psi_R(x_1)\overline{\psi}_R(x_2) \psi_L(x_2):   \non \\[2mm]
& + & :\overline{\psi}_R(x_1) \psi_L(x_1)\overline{\psi}_L(x_2) \psi_R(x_2):+:\overline{\psi}_R(x_1) \psi_L(x_1)\overline{\psi}_R(x_2) \psi_L(x_2): \non \\[2mm]
&+&  2 \,  i \lf(S^L_{\al \bt}(x_2-x_1) \, :\overline{\psi}^\bt_R(x_2) \psi^\al_R(x_1):+ S^R_{\al \bt}(x_2-x_1) \, :\overline{\psi}^\bt_L(x_2) \psi^\al_L(x_1):\ri) \Big] \, , \label{secu}
\eea
where
\be
S^\si_{\al \bt}(x) \ = \ \int \!\! \frac{\dr^4 p}{(2 \pi)^4} \, e^{-i p x} \, \frac{P_\si \, \slashed{p}_{\al \bt}}{p^2+i \varepsilon} \, , \qquad \si=L,R \, ,
\ee
is the Dirac propagator for free fields. The corresponding diagrams are depicted in Fig. \ref{fig:2th}. The divergent vacuum contributions need to be subtracted, see the discussion in Ref. \cite{Blasone:2023brf}.

The survival process is
\be
|\psi_{\G p,L}\ran \ \rightarrow \ |\psi_{\G p,L}\ran \, .
\ee
Maintaining only up to quadratic terms in $m$, the survival amplitude can be written as
\be
\mathcal{A}_{L \to L}(\G p; t_i,t_f) \ = \ 1 -\ha \mathcal{A}^{(2)}_{L \to L}(\G p; t_i,t_f) \, , 
\ee
where $A^{(2)}_{L \to L}(\G p;t_i,t_f)$ is second order contribution (proportional to $m^2$). Taking the square and disregarding all terms with power of the mass larger than 2, we find
\be
\mathcal{P}_{L \to L}(\G p; \De t) \ = \  1-  \, \Re e \lf(\mathcal{A}^{(2)}_{L \to L}(\G p;t_i,t_f)\ri) \, . 
\ee
The amplitude explicitly reads
\bea \non
 \mathcal{A}^{(2)}_{L \to L}(\G p;t_i,t_f)  \approx
  m^2 \,  \int^{t_f}_{t_i} \!\! \dr t_1  \,  \int^{t_f}_{t_i} \!\! \dr t_2 \, e^{2 i |\G p|(t_1-t_2)}   = \  \frac{m^2}{p^2}  \,  \, \sin^2 (|\G p|(t_f-t_i)) \, .
\eea
Then the result is
\be \label{perfor2}
\mathcal{P}_{L \to L}(\G p; \De t) \ = \   1-  \frac{m^2}{|\G p|^2}  \sin^2\lf(|\G p| \, \De t\ri) \, . 
\ee
At the lowest order this coincides with Eq. \eqref{plr}. In fact
\be
\frac{m^2}{\om_\G p^2}  \sin^2\lf(\om_\G p \, \De t\ri) \ = \ \frac{m^2}{|\G p|^2}  \sin^2\lf(|\G p| \, \De t\ri) +O(m^4) \, .
\ee
We thus recovered the quantum mechanical result employing QFT at finite time. 

\begin{figure}[t]
\centering
\includegraphics[scale=0.6]{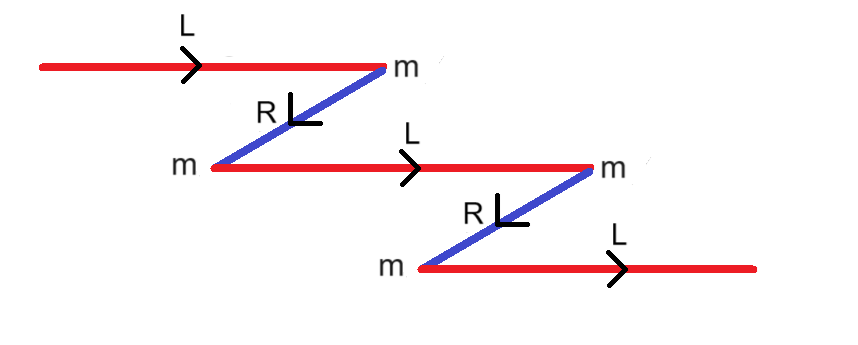}
\caption{Fourth-order diagram contributing to the $L$ survival probability. }
\label{fig:4th}
\end{figure}

Although the above is an approximate result, it is evident that the coincidence of Eq.\eqref{perfor2} with  Eq.\eqref{plr} is highly non-trivial. In order to understand how to go beyond the leading order, we look at the fourth order. A contribution proportional to $m^4$ is provided by the square of the second-order amplitude 
\be
\frac{1}{4}|\mathcal{A}^{(2)}_{L \to L}(\G p; t_i,t_f)|^2 \ = \ \frac{m^4}{4 |\G p| ^4} \sin^4(|\G p|  \De t) \, .
\ee
However, this is not the only term proportional to $m^4$. From the fourth-order amplitude, see Fig. \ref{fig:4th}, we get
\be
\frac{1}{24} \mathcal{A}^{(4)}_{L \to L}(\G p; t_i,t_f) \ = \ \frac{m^4}{24 |\G p| ^4} \sin^4(|\G p|  \De t) \, .
\ee 
Therefore
\be \label{perfor4}
\mathcal{P}_{L \to L}(\G p; \De t) \ = \   1-  \frac{m^2}{|\G p|^2}  \sin^2\lf(|\G p| \, \De t\ri)+\frac{ m^4}{3 |\G p| ^4} \sin^4(|\G p|  \De t)  \, . 
\ee
In order to check whether this is consistent with the exact survival probability of Eq. (\ref{plr}), we expand the latter up to the fourth order in $m$, obtaining
\be
\mathcal{P}_{L \to L}(\G p;  \De t) \ \approx \ 1-\frac{m^2 \sin ^2(|\G p| \De t)}{|\G p|^2}-\frac{m^4 \sin (|\G p| \De t) (|\G p| \De t \cos (|\G p| \De t)-\sin (|\G p| \De t))}{|\G p|^4} \, .
\ee
Clearly, such expansion is only valid at short time intervals: for large $\De t$, the last piece would grow indefinitely. Further expanding the expression in $\De t$, we get
\be
\mathcal{P}_{L \to L}(\G p; \De t) \ \approx \ 1-m^2 \De t^2+\frac{1}{3} m^2 \De t^4 \om^2_{\G p}+O\left(\De t^5\right) \, ,
\ee
where $\om_{\G p}=\sqrt{|\G p|^2+m^2}$ is the energy of the massive fermion.
It is easy to see that this is the same result we obtain from Eq.\eqref{perfor4} expanding in $\De t$ and taking into account that $ \sin\lf(|\G p| \, \De t\ri) \approx |\G p| \De t$.

\begin{figure}[t]
\centering
\includegraphics[scale=0.6]{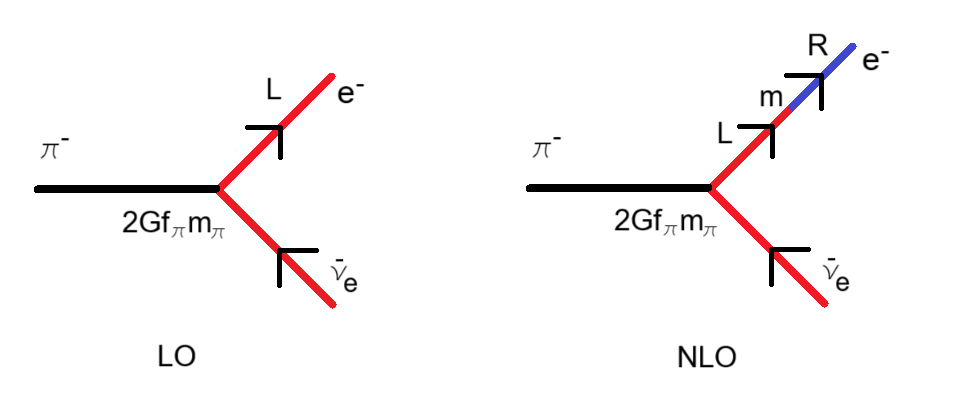}
\caption{LO and NLO diagrams of $\pi^-$ into
$e^- \bar{\nu}_e$. LO vanishes because of spin conservation. NLO is nonzero and well approximates the full result.
}
\label{fig:piondecay}
\end{figure}

Let us now consider the decay of a pion, $\pi^- \to e^- + {\bar \nu}_e$, see Fig. \ref{fig:piondecay}. For simplicity, we assume the antineutrino to be exactly massless. In the rest frame of the pion, the two leptons are emitted with opposite momenta along the $z$-axis; we denote the outgoing electron momentum with
$\mathbf{k}_{1}=\mathbf{k=}(0,0,k_{z}>0)$ and the neutrino one with
$\mathbf{k}_{2}=-\mathbf{k=}(0,0,-k_{z}<0)$.
To conserve angular momentum, the leptons must also have opposite helicities. Since the antineutrino is massless, it can only be produced with negative helicity in a weak decay. Consequently, the electron must have positive helicity. This is only possible if the electron mass is nonzero, as weak interactions produce left-chiral leptons.
The effective Lagrangian for the pion decay reads:%
\begin{equation}
\mathcal{L}_{\pi}=2Gf_{\pi}\partial_{\mu}\pi^{-}\bar{\psi}_{e,L}\gamma^{\mu
}\psi_{\nu_{e},L}+h.c.
\end{equation}
with the Fermi constant $G=g_{w}^{2}/8M_{W}^{2}$ and the pion decay constant
$f_{\pi}=92.4$ MeV \cite{ParticleDataGroup:2024cfk}.
Treating the mass term as a
perturbation and keeping the fermions as massless, the LO decay amplitude (left part of Fig. \ref{fig:piondecay})
vanishes exactly:
\begin{equation}
i\mathcal{A}_{LO}=i2Gm_{\pi}f_{\pi}\bar{u}_{L}(k_{z})\gamma^{0}%
v_{L}(-k_{z})=0\text{ ,}%
\end{equation}
where the spinor $u$ refers to the electron and $v$ to the neutrino, see Fig. \ref{fig:piondecay}.

At NLO (right part of Fig. \ref{fig:piondecay}) the mass induces a chiral oscillation ($L  \rightarrow R$) that allows for a nonzero decay amplitude: 
\begin{equation}
i\mathcal{A}_{NLO}=i2Gm_{\pi}f_{\pi}m\left(  \bar{u}_{R}(k_{z})P_{L}%
\frac{\not k_{1}}{k^{2}}\gamma^{0}v_{L}(-k_{z})\right)  =4Gm_{\pi}f_{\pi
}mk_{z}\frac{\omega_{\bold{k}}-k_{z}}{k^{2}}=2Gf_{\pi}m
\text{ ,}
\label{nlo}
\end{equation}
where on the r.h.s. a fictitious infinitesimal mass $\delta$ for the electron 
\begin{equation}
\frac{\omega_{\mathbf{k}}-k_{z}}{k^{2}}\simeq\frac{k_{z}+\frac{\delta^{2}%
}{2k_{1,z}}-k_{1,z}}{\delta^{2}}=\frac{1}{2k_{z}}%
\end{equation}
has been used. In this limit, the width is
\begin{equation}
\Gamma_{\pi^{-}\rightarrow e^{-}\bar{\nu}_{e}}=k_{z}\frac{G^{2}f_{\pi}%
^{2}m^{2}}{2\pi}=\frac{G^{2}f_{\pi}^{2}m^{2}m_{\pi}}{4\pi}
\text{ ,}
\end{equation}
which for small $m$ is in agreement with the full result $\Gamma_{\pi
^{-}\rightarrow e^{-}\bar{\nu}_{e}}=\frac{G^{2}}{4\pi}f_{\pi}^{2}m_{\pi}%
m^{2}\left(  1-m^{2}/m_{\pi}^{2}\right)  ^{2}$, see e.g. Ref. \cite{Greiner:1993qp}.

Quite remarkably, the full decay width can be obtained within our
\textquotedblleft perturbative \textquotedblright\ approach. Basically, the NLO,
NNNLO,...diagrams can be obtained by considering the full NLO expression of
Eq. (\ref{nlo}) and by replacing $\delta^{2}\rightarrow m^{2}$. Thus, the mass
$m$ appears both as a vertex and within the kinematic terms\footnote{The
resummed amplitude NLO+NNNLO+... reads $i\mathcal{A}_{res}=2Gm_{\pi}f_{\pi
}m\frac{\sqrt{\omega_{\mathbf{k}}+m}+k_{z}}{\sqrt{\omega_{\mathbf{k}}+m}}%
\sqrt{k_{z}}\frac{\omega_{\mathbf{k}}-k_{z}}{m^{2}}$. This leads to the exact result for the decay width.
Note, the amplitude above coincides with the one obtained by considering the left-handed electron with `inverted' positive helicity: $2Gm_{\pi}f_{\pi}%
\bar{u}_{L}^{(\lambda=+1)}(k_{1,z})\gamma^{0}v_{L}(-k_{1,z})$.
%that leads to resulting amplitude $2Gm_{\pi}f_{\pi}\frac{E_{1}+m-k_{z}}{\sqrt{E_{1}+m}}%
%\sqrt{k_{z}}$ coincides with the one above. 
}. 
This approach corresponds to
the resummation of the external fermion line.
Summarizing, the decay of the pion is a neat example of a virtual chiral oscillation. 

%\iffalse
%***
%the mass is effectively introduced as an interaction vertex. A virtual electron with negative helicity and left-chirality is initially produced, but due to the mass vertex,  chirality oscillates, thus making the decay possible. 
%****
%
%The equivalence of these two descriptions can be verified by analyzing the leptonic part of the decay amplitude
%\be \label{nppd}
%\overline{u}_{k_z,L}^+ \ga_0 v_{-k_z, L} \ = \ \frac{(\om_k+k_z) (\om_k-k_z+m)}{2 \sqrt{\om_k (\om_k+m)}} \, , 
%\ee
%where $\overline{u}_{L,k_z}^+$ is the massive electron spinor with left chirality and positive helicity, while $v_{L,-k_z}$ denotes the spinor for the antineutrino. From a perturbative perspective, if we treat the electron as massless, it is straightforward to verify that the leading-order result is $\overline{u}_{L,k_z} \ga_0 v_{L,-k_z}=0$
%. This result arises because a massless electron cannot simultaneously possess positive helicity and left chirality.
%To obtain a nontrivial result for comparison, we must move to the next-to-leading order, specifically the linear order in the electron mass $m$:
%\be \label{ppd}
%\lim_{\varepsilon \to 0} \, m \, \overline{u}_{k_z, R} P_L \frac{\om_k \ga_0-k_z \ga_3}{\om_k^2-k_3^2+i \varepsilon} \ga_0 v_{-k_z, L} \ = \ m \, , 
%\ee
%where now all spinors refer to massless particles. The main observation is that the expressions in \eqref{nppd} and \eqref{ppd} coincide at the linear order in $m$. The corresponding diagram is shown in the figure. In the perturbative framework, pion decay is allowed due to a virtual chiral oscillation of the electron. 
%\fi 

The pion case raises the question: is it possible to detect the real right-handed electron in weak interactions? 
This is feasible only if chiral oscillations occur. Let us consider, e.g. the detection process $e^- + X \to \nu_e+ Y$, where an initially left-handed electron scatters with a nucleus $X$, producing a neutron (within the new nucleus $Y$) and an electron neutrino. 
In particular, the effective interaction is proportional to $J_{\mu}\bar{\psi
}_{e,L}\gamma^{\mu}\psi_{\nu_{e},L}$ where $J_{\mu}$ is the current that
describes the nucleus states $X$ and $Y.$ The ingoing electron has momentum
$\mathbf{p}=(0,0,p_{z}>0),$ the outgoing neutrino $\mathbf{k.}$ The leptonic
part of the amplitude is proportional to $\bar{u}_{L}(\mathbf{k})\gamma^{\mu
}u_{L}(\mathbf{p}).$ Assuming an \textquotedblleft heavy\textquotedblright%
\ target such that the $\mu=0$ component dominates, for a scattering angle
$\theta\approx0$ the leptonic amplitude reduces to $u_{L}^{\dagger}%
(\mathbf{p})u_{L}(\mathbf{p})\simeq 2p_{z}$. Thus, the differential
cross-section $d\sigma_{L}/d\Omega(\theta\approx0)$ is expected to be sizable.

However, if the left-handed neutrino oscillates into a right-handed
one, the very same amplitude vanishes at LO. It is nonzero at the NLO because
the right-handed neutrino oscillates back into a left-handed one. Yet, the
amplitude is suppressed in a way that resembles the pion decay. Summarizing,
the occurrence of chiral oscillations reduces the differential small-angle cross-section.
A thorough study of scattering off nuclei that involves realistic initial
states and the whole kinematical analysis is an interesting extension, constituting a promising outlook.

\iffalse
Let us assume the proton is at rest and that the electron moves along the $z$-axis. We can then take as plane for the neutrino motion the $x-z$ plane. The lepton current to be evaluated is now $\overline{u}_{\G p,L} \ga_\mu u^+_{k_z,L}$, where $u^+_{k_z,L}$ is the electron spinor, while $\overline{u}_{\G p,L}$ is the neutrino one, with $\G p=(p_x,0,p_z)$. We can compute, e.g. the 0 component, obtaining
\be \label{npdec}
\overline{u}_{\G p,L} \ga_0 u^+_{k_z,L} \ = \ -\frac{p_x \left(\om_k-k_z+m\right)}{2 \sqrt{\om_k+m} \sqrt{|\G p|}}
\ee
As in the case of pion decay, the analogous leading-order expression in the perturbative framework (i.e., the corresponding expression for a massless electron) evaluates to zero. However, the next-to-the-leading-order expression gives
\be
\lim_{\varepsilon \to 0} \, m \, \overline{u}_{\G p,L} \ga_0 P_L \frac{\om_k \ga_0-k_z \ga_3}{\om_k^2-k_3^2+i \varepsilon}  u_{k_z,R} \ = \ \frac{m \, p_x}{2 \, \sqrt{k_z \, |\G p| }}
\ee
which coincides with Eq.\eqref{npdec} at the linear order in $m$. The corresponding diagram is show in figure.
\fi
%%%%%%%%%%%%%%%%%%%%%%%%%%%%%%%%%%%%%%%%%%%%%%%%%%%%%
\section{Conclusions} \label{con}
In this paper we have demonstrated that chiral oscillations can be studied in perturbative QFT employing the usual Dyson series for the time evolution operator in the interaction picture. In the present case, in fact, the mass term of the Dirac Lagrangian is regarded as a coupling between massless fields with (opposite) definite chiralities. Within the approximation adopted in the perturbative expansion, our result non-trivially coincides with the usual quantum mechanical oscillation formula \eqref{plr} reported in  Ref. \cite{Bittencourt:2024yxi}.
The interaction picture computation presented in this paper provides the chiral oscillation formula up to fourth order in the interaction coupling $m$. Given that the interaction here (and in Ref. \cite{Blasone:2023brf} concerning flavor oscillations) is quadratic in the fermion fields, it is plausible that an exact, albeit complex, resummation is achievable. This possibility will be explored in future research.

Furthermore, we have shown that the perturbative framework for chiral oscillations provides an effective description of production processes in which an electron must simultaneously possess left chirality and positive helicity, as in the decay $\pi^- \to e^- + {\bar \nu}_e$. 
It is worth stressing that, while the chiral oscillation formula concerns the time evolution of real (on-shell) particles, the chiral oscillation involved in the perturbative view of the pion decay, involves virtual (off-shell) electrons.
Finally, we argue that, from this perturbative perspective, chiral oscillations are also essential for the evaluation of scattering processes aiming at their detection.

%Another important aspect to be stressed is that the results of the present analysis are in agreement with those of Ref.\cite{Bittencourt:2024yxi}, which were derived within an   algebraic approach, based on the diagonalization of the chiral charge operators. 

%%%%%%%%%%%%%%%%%%%%%%%%%%%%%%%%%%%%%%%%%%%%%%%%%%%%%
\section*{Acknowledgments}
G.T. would like to thank FAPESP  2023/06278-2 and CNPq bolsa de produtividade 305731/2023-8 for the financial support.
%%%%%%%%%%%%%%%%%%%%%%%%%%%%%%%%%%%%%%%%%%%%%%%%%%%%%%%%%%%%%%%%%%%%%
%\appendix
%
%%%%%%%%%%%%%%%%%%%%%%%%%%%%%%%%%%%%%%%%%%%%%%%%%%%%%%%%%%%%%%%%%%%%%%%
%%%%%%%%%%%%%%%%%%%%%%%%%%%%%%%%%%%%%%%%%%%%%%%%%%%%%%%%%%%%%%%%%%%%%%%%
\bibliography{LibraryNeutrino}

\bibliographystyle{apsrev4-2}

\end{document}